\documentclass[12pt]{article} 

\usepackage[utf8]{inputenc}
\usepackage[T1]{fontenc} 
\usepackage[english]{babel}           

\usepackage[pdftex]{graphicx}
\usepackage{epsfig}
\usepackage{graphicx}
\usepackage{comment}
\usepackage{latexsym}
\usepackage{hyperref}
\usepackage{amsmath}
\usepackage{mathrsfs}
\usepackage[usenames, dvipsnames]{color}
\usepackage{amsbsy}
\usepackage{amssymb}
\usepackage{amsthm}
\usepackage{amsfonts}
\usepackage{cite}
\usepackage{enumitem}
\usepackage{xcolor}
\usepackage{color}
\usepackage{mathtools}
\usepackage{caption} 
\usepackage{subcaption} 

\usepackage{diagbox}

\newcommand{\be}[0]{\begin{equation}}
\newcommand{\ee}[0]{\end{equation}}

\renewcommand{\thefootnote}{\fnsymbol{footnote}}

\newcommand{\Z}{\mathbb{Z}}

\renewcommand{\O}{{\cal O}}

\renewcommand{\and}{\mbox{and}}

\newcommand{\bm}{\boldmath} 

\newcommand{\N}{{\cal N}}

\newcommand{\Msusy}{M_{\rm susy}}
\newcommand{\Ms}{M_{\rm string}}
\newcommand{\Mc}{M_{\rm c}}

\catcode`\@=11
\def\marginnote#1{}
\newcount\hour
\newcount\minute
\newtoks\amorpm
\hour=\time\divide\hour by60 \minute=\time{\multiply\hour by60
\global\advance\minute by-\hour}
\edef\standardtime{{\ifnum\hour<12 \global\amorpm={am}%
        \else\global\amorpm={pm}\advance\hour by-12 \fi
        \ifnum\hour=0 \hour=12 \fi
        \number\hour:\ifnum\minute<10 0\fi\number\minute\the\amorpm}}
\edef\militarytime{\number\hour:\ifnum\minute<10 0\fi\number\minute}
\def\draftlabel#1{{\@bsphack\if@filesw {\let\thepage\relax
   \xdef\@gtempa{\write\@auxout{\string
      \newlabel{#1}{{\@currentlabel}{\thepage}}}}}\@gtempa
   \if@nobreak \ifvmode\nobreak\fi\fi\fi\@esphack}
        \gdef\@eqnlabel{#1}}
\def\@eqnlabel{}
\def\@vacuum{}
\def\draftmarginnote#1{\marginpar{\raggedright\scriptsize\tt#1}}
\def\draft{\oddsidemargin -.2truein
        \def\@oddfoot{\sl preliminary draft \hfil
        \rm\thepage\hfil\sl\today\quad\militarytime}
        \let\@evenfoot\@oddfoot \overfullrule 3pt
        \let\label=\draftlabel
        \let\marginnote=\draftmarginnote
   \def\@eqnnum{(\theequation)\rlap{\kern\marginparsep\tt\@eqnlabel}%
\global\let\@eqnlabel\@vacuum}  }
\def\thebibliography#1{
\vskip 0.5cm \centerline{\bf \Large References}
\list{
[\arabic{enumi}]}{\settowidth\labelwidth{[#1]}
\leftmargin\labelwidth
\advance\leftmargin\labelsep
\usecounter{enumi}}
\def\newblock{\hskip .11em plus .33em minus .07em}
\sloppy\clubpenalty4000\widowpenalty4000
\sfcode`\.=1000\relax}

\renewcommand{\theequation}{\arabic{section}.\arabic{equation}}
\renewcommand{\section}{\setcounter{equation}{0}\@startsection
{section}{1}{0mm}{-\baselineskip}{0.5\baselineskip} {\normalfont\Large\bfseries}}
\renewcommand{\subsection}{\@startsection
{subsection}{2}{0mm}{-\baselineskip}{0.5\baselineskip} {\normalfont\large\bfseries}}
\renewcommand{\subsubsection}{\@startsection
{subsubsection}{3}{0mm}{-\baselineskip}{0.5\baselineskip}
{\normalfont\normalsize\slshape}}

\begin{document}


\begin{titlepage}
\begin{flushright}
CPHT-RR066.112019, November 2019
\vspace{1.5cm}
\end{flushright}
\begin{centering}
{\bm\bf \Large PHASE TRANSITION AT HIGH SUPERSYMMETRY \\ 
\vspace{.4cm} BREAKING SCALE IN STRING THEORY}

\vspace{9mm}

 {\bf Herv\'e Partouche and Balthazar de Vaulchier}

 \vspace{7mm}

{CPHT, CNRS, Ecole polytechnique, IP Paris, \\F-91128 Palaiseau, France\\ \textit{herve.partouche@polytechnique.edu, balthazar.devaulchier@polytechnique.edu}}

\end{centering}
\vspace{0.3cm}
$~$\\
\centerline{\bf\Large Abstract}\\
\vspace{-1cm}

\begin{quote}

\hspace{.2cm} 

When supersymmetry is spontaneously broken at tree level, the spectrum of the heterotic string compactified on orbifolds of tori contains an infinite number of potentially tachyonic modes. We show that this implies instabilities of Minkowski spacetime, when the scale of supersymmetry breaking is of the order of the string scale. We derive the phase space structure of vacua in the case where the tachyonic spectrum contains a mode with trivial momenta and winding numbers along the internal directions not involved in the supersymmetry breaking.

\end{quote}

\end{titlepage}
\newpage
\setcounter{footnote}{0}
\renewcommand{\thefootnote}{\arabic{footnote}}
 \setlength{\baselineskip}{.7cm} \setlength{\parskip}{.2cm}

\setcounter{section}{0}


\section{Introduction}
\label{intro}

Phase transitions occur in various contexts in high energy physics. The most common setup describing such effects is the Brout-Englert-Higgs mechanism, which  occurs when a scalar field $\phi$ becomes tachyonic. When the squared mass is negative, $\phi$ sits  at a maximum of the scalar potential and therefore condenses. The  new vacuum expectation value (vev) of $\phi$ minimizes  (locally) the potential, and the theory has switched from a ``wrong'' to a ``true'' vacuum.  
What we review in the present note is that a similar condensation occurs in string theory, when the scale  $\Msusy$ of spontaneous supersymmetry breaking is of the order of the string scale $\Ms$~\cite{PV}. 

To be specific, we consider classical string models in Minkowski spacetime, where supersymmetry is spontaneously broken. Because there is only one true constant scale in the theory, which is $\Ms$, the scale $\Msusy$ is a field the tree level potential $V$ depends on. Our assumption on flatness of the classical background amounts to saying that minima of $V$ lie at $V=0$. It turns out that local supersymmetry implies the latter to be degenerate, and that one of the flat directions is parameterized by the field $\Msusy$ itself. For this reason, the supergravity models describing the spontaneous breaking of supersymmetry in flat space are referred as ``no-scale models''~\cite{noscale}, since there is no preferred value for the vev $\langle \Msusy\rangle$ at tree level. In the framework of string theory, this statement is actually valid up to a critical value $\Mc$ of $\langle \Msusy\rangle$, which is of the order of $\Ms$. Above this bound, the condensation of a tachyonic scalar triggers a second order phase transition from the no-scale phase to a new phase, which is argued to be associated with a non-critical string theory. Even though this phenomenon is physically very different from the Hagedorn phase transition encountered in string theory at finite temperature $T$, when the latter is of the order of $\Ms$~\cite{Atick}, it turns out to be similar from a technical point of view~\cite{AK, ADK}.  

In its usual formulation, string theory is defined in first quantized formalism. This means that what is known (at least in principle) is the massless and massive spectrum that is allowed to populate a consistent vacuum described by a  conformal field theory on the worldsheet. In order to find the shape of the potential far from the vacuum under consideration, one should in principle evaluate an infinite number of correlation functions, and resum them in order to reconstruct  the full expression of the off-shell tree level potential. Alternatively, we may consider in principle a second quantized formulation of string theory, i.e. string fields theory, in order to derive the potential. However, given the fact that we are only interested in the vacuum structure of the tree level potential, we will analyze the problem at low energy, in the effective supergravity. 

In Sect.~\ref{II}, we introduce a class of string theory no-scale models in four dimensions that realize the $\N=4\to \N=0$ spontaneous breaking of supersymmetry. 
In Sect.~\ref{III}, we implement an orbifold action that reduces the initial $\N=4$ supersymmetry  to  $\N=1$, and we present the  necessary ingredients  to derive the tree level potential $V$ in presence of super-Higgs mechanism. The final expression of $V$  is presented in  Sect.~\ref{IV}, where the different phases of the theory are derived. Our conclusions can be found in Sect.~\ref{V}. 

\section{$\N=4\to \N=0$ Heterotic No-Scale Models}
\label{II}

Our starting point is the heterotic string compactified on a 6-dimensional torus, where supersymmetry is spontaneously broken by a stringy version~\cite{SSstring4} of the Scherk-Schwarz mechanism~\cite{SS2,PZ}. In field theory, the latter is a refined version of the Kaluza-Klein reduction we first present in its simplest possible realization. Let us consider a field theory in $4+1$ dimensions, where the extra coordinate is compactified on a circle of radius $R_4$. Assuming the existence of a symmetry with conserved charge $Q$ in $4+1$ dimensions, we may impose $Q$-dependent boundary conditions for every field $\varphi$, which translate into Kaluza-Klein masses $M$ for its Fourrier modes $m_4\in\Z$, 
\begin{equation}
\varphi(x^\mu,x^4) = {1\over\sqrt{2\pi R^4}}\sum_m\varphi_m(x^\mu)\, e^{i{m_4+e Q\over R_4}x^4}\;\;\Longrightarrow \;\; M^2 = \left({m_4+eQ\over R_4}\right)^2\, .
\end{equation}
In the above formulas, $\mu\in\{0,\dots,3\}$ and we have included a parameter $e=1$ or 0 in order to describe both Scherk-Schwarz and Kaluza-Klein cases, respectively. When the higher dimensional theory is supersymmetric and we choose $Q\equiv {F\over 2}+Q_{\rm susy}$, where $F$ is the fermionic number and $Q_{\rm susy}$ is a constant charge  within each supermultiplet, the boson/fermion degeneracy in four dimensions is lifted and the theory describes a super-Higgs mechanism, with scale $\Msusy=e/(2R_4)$. 

In the $E_8\times E_8$ heterotic string compactified on a factorized torus $T^6\equiv S^1(R_4)\times T^5$, the previous mass formula in string units ($\Ms=1$) is generalized to ~\cite{SSstring4} 
\begin{equation}
\label{m2}
M^2=\left({m_4+eQ-{n_4\over 2}e^2\over R_4}+n_4 R_4\right)^2 +2\Big[( Q-en_4)^2+Q_2^{2}+Q_3^{2}+Q_4^{2}-1\Big],
\end{equation}
where $n_4\in \Z$ is the winding number of the string along $S^1(R_4)$, and $\vec Q\equiv (Q,Q_2,Q_3,Q_4)$ is a quadruple of charges arising from the fact that for $\mbox{$e=0$}$ the theory is $\N=4$ supersymmetric. The above equation applies to the lightest modes, which in the bosonic sector have $(Q,Q_2,Q_3,Q_4)=(\pm1,0,0,0)$ or permutations. Notice the presence of the $-1$ contribution in the squared brackets, which is the zero point energy arising from the quantization of the fields on the worldsheet. In the supersymmetric case ($e=0$), we have $M^2\ge 0$ for all modes, while in the spontaneously broken case ($e=1$), the dangerous contribution $-1$ is not canceled when $Q=n_4=\pm 1$. Looking at this fact more closely, one finds that the pair of scalar states $m_4=-n_4=-Q=\epsilon$, where $\epsilon=\pm 1$, are tachyonic when 
\begin{equation}
\label{range}
{\sqrt{2}-1\over \sqrt{2}}\equiv{1\over 2R_c}<R_4<R_c\equiv{\sqrt{2}+1\over \sqrt{2}}\, .
\end{equation}  
Therefore, an instability arises in the theory when the supersymmetry breaking scale $\Msusy$ reaches the critical value $\Mc=1/(2R_c)$. 

\begin{figure}[h]
\begin{center}
\includegraphics[scale=1]{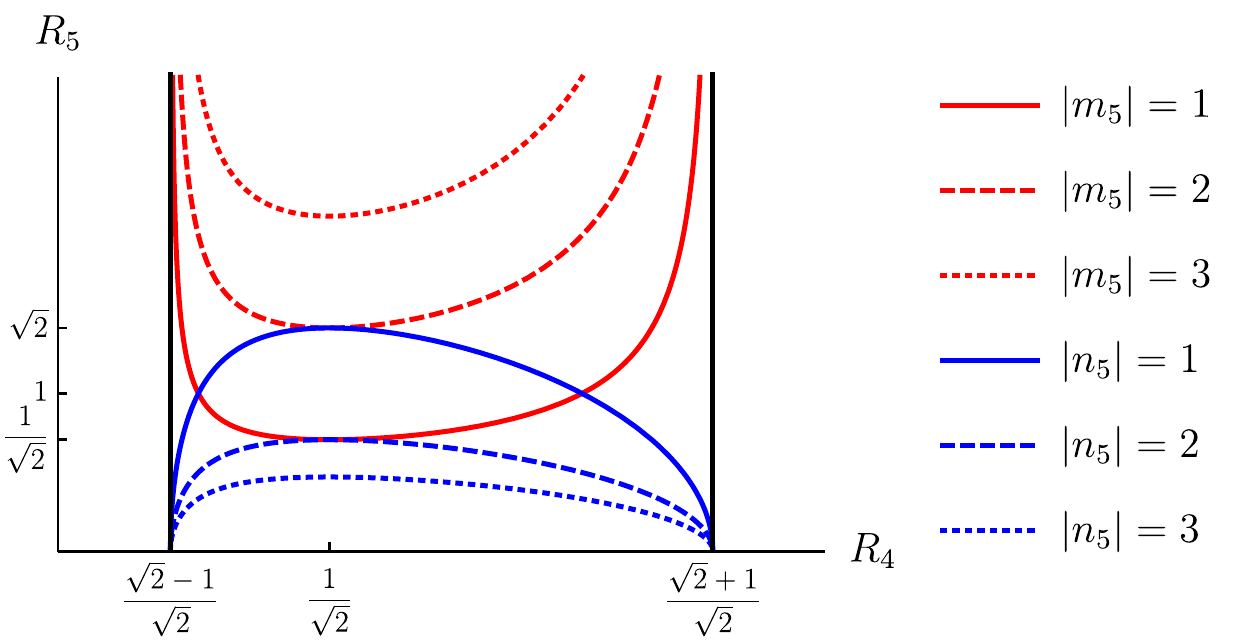}
\end{center}
%
%
\caption{Boundary curves of the regions of the plan $(R_4,R_5)$, where Kaluza-Klein or winding modes along $S^1(R_5)$ are tachyonic.}
\label{fig1}       
\end{figure}

Moreover, taking into account the fact that the tachyonic modes may also have non-trivial momentum $m_5\in \Z$ or winding number $n_5\in \Z$ (but not both, due to the left/right-level matching) along one more internal direction $X^5$, their mass formula becomes 
\begin{equation}
\label{mass}
M^2={1\over 4R_4^2}+R_4^2-3+\Big({m_5\over R_5} \Big)^2\quad \mbox{or}\quad  M^2={1\over 4R_4^2}+R_4^2-3+(n_5 R_5 )^2\, , 
\end{equation}
where we have assumed for simplicity the  internal space to be factorized as $T^6\equiv S^1(R_4)\times S^1(R_5)\times T^4$. Therefore, the larger (smaller) $R_5$ is, the larger the number of tachyonic momentum (winding) states along $S^1(R_5)$ is, as shown in Fig.~\ref{fig1}. One of our  goal is then to see  whether the infinity of potentially tachyonic modes yield a multiphase diagram or not, beside the no-scale-phase we started with. Of course, even if we will not do so, this question may be considered in the most general case, where the momenta and winding numbers along the remaining internal radii directions of $T^4$ are taken into account. 

Before concluding this section, let us specify what conserved charges $Q$ may be used to implement the $\N=4\to 0$ Scherk-Schwarz breaking of supersymmetry. On the left-moving supersymmetric side of the heterotic string, we can rotate any pair of worldsheet fermions $\psi^a,\psi^b$,  where $a,b\in\{2,\dots,9\}$ in light cone gauge. The charges $Q$ are then the eigenvalues of the generator associated with one of the $O(2)$ affine algebra currents $:\psi^a\psi^b\!:$. Because all $\psi^a$'s have identical boundary conditions on the worldsheet, all pairs $(a,b)$ yield equivalent non-supersymmetric models when $e=1$. 


\section{Gauged $\N=4$ Supergravity Truncated to $\N=1$}
\label{III}

Gauged $\N=4$ supergravity contains a gravity multiplet coupled to an arbitrary number $6+k$ of vector multiplets~\cite{sugra2, sugra3, sugra4,sugra2bis, sugraN=4}. The scalar content is a complex field $\Phi$ and $6\times (6+k)$ real scalars $Z_a^S$, $a\in\{4,\dots,9\}$, $S\in\{4,\dots,15+k\}$, defining a non-linear $\sigma$-model with target space
\begin{equation}
{SU(1,1)\over U(1)}\times {SO(6,6+k)\over SO(6)\times SO(6+k)}\, .
\end{equation}
The coordinates of the second coset satisfy $\eta_{ST} Z_a^SZ_b^T=-\delta_{ab}$, where  $\eta=\mbox{diag}(-1,\dots,-1,1,\dots)$ with 6 entries $-1$. 
To diminish the number of degrees of freedom and simplify the analysis, we implement from now on a $\Z_2\times \Z_2$ orbifold action on the parent supersymmetric heterotic model, which reduces $\N=4$ to $\N=1$. The generators $G_1,G_2$ act respectively as twists $X^a\to -X^a$ on the directions $X^6,X^7,X^8,X^9$ and $X^4,X^5,X^8,X^9$, thus reducing $T^6$ to $ S^1(R_4)\times S^1(R_5)\times T^2\times T^2$. In that case, the choice of charge $Q$ must be  compatible with the orbifold action. A consistent choice amounts to taking the $O(2)$ current with $a=6, b=8$ (i.e. in {\em distinct} $T^2$ tori). To convince ourselves, let us note that the tachyonic modes, say with pure momenta along $T^2\times T^2$, transform consistently into each other under $G_1$ and $G_2$:\footnote{In our notations, $e^{ip_{I\rm L}X^I_{\rm L}+ip_{I\rm R}X^I_{\rm R}}|0\rangle_{\rm NS}\otimes |\tilde 0\rangle$ stands for $|p_{\rm L}\rangle_{\rm NS}\otimes |\widetilde{p_{\rm R}}\rangle$, where the coordinates and the generalized momenta are divided into their left- and right-moving contributions, $X^I=X^I_{\rm L}+X^I_{\rm R}$, $p_I=p_{I\rm L}+p_{I\rm R}$, and where we have set $R_4=1/\sqrt{2}$ for convenience. }  
\begin{eqnarray}
&&\phantom{\longrightarrow-(-1)^\xi\,  }{\psi^6+i\epsilon \psi^8\over \sqrt{2}}\, e^{i\epsilon X^4_{\rm R}}\,e^{ip_{5\rm L}X^5}\, e^{i\sum_{I=6}^9 p_{I\rm L}X^I} |0\rangle_{\rm NS}\otimes |\tilde 0\rangle\nonumber \\
&&\longrightarrow-(-1)^\xi\, {\psi^6+i\epsilon \psi^8\over \sqrt{2}}\, e^{i\epsilon X^4_{\rm R}}\,e^{ip_{5\rm L}X^5}\, e^{-i\sum_{I=6}^9 p_{I\rm L}X^I} |0\rangle_{\rm NS}\otimes |\tilde 0\rangle\label{orbifold}
\\
&&\longrightarrow{\psi^6-i\epsilon \psi^8\over \sqrt{2}}\, e^{-i\epsilon X^4_{\rm R}}\,e^{-ip_{5\rm L}X^5}\, e^{i\left(p_{6\rm L}X^6+p_{7\rm L}X^7-p_{8\rm L}X^8-p_{9\rm L}X^9\right)} |0\rangle_{\rm NS}\otimes |\tilde 0\rangle.\nonumber 
\end{eqnarray}
On the contrary, with $(a,b)=(6,7)$ or $(8,9)$, the generator $G_1$ would inconsistently send the tachyons into massive superpartners. In the above formula, we have introduced a discrete torsion $\xi=1$ or 0 that yields two drastically different patterns of  tachyonic modes surviving the $G_1$-orbifold action.\footnote{See the revised version of arXiv:1903.09116 \cite{PV}.} In the following, we restrict ourselves to the analysis of the case $\xi=1$. Notice that since the $O(2)$ generator used to implement the Scherk-Schwarz breaking of $\N=1$ supersymmetry rotates directions of distinct $T^2$'s, some of the tori deformation moduli are projected out. 

Our goal is to derive the $\N=1$ supergravity potential $V$ that depends on the scalar fields whose masses are given in Eq.~(\ref{mass}), and on the radii $R_4,R_5$ and the dilaton field. This amounts to freezing (artificially) all remaining moduli, which are associated with $(i)$ the internal $T^2\times T^2\times T^2$ $(ii)$ or $E_8\times E_8$ Wilson lines, $(iii)$ or which arise from the  twisted sectors. Moreover, as said before, we do not include the potentially tachyonic modes with non-trivial momentum or winding numbers along $X^6,X^7,X^8,X^9$, which we expect would not change the final phase diagram for the choice of discrete torsion $\xi=1$ considered in this work. In that case, we find convenient to derive the result by truncating suitably the $\N=4$ gauged supergravity associated with the parent $\N=4\to \N=0$ heterotic no-scale model. The non-linear $\sigma$-model reduces to 
\begin{equation}
\label{cosets}
{SU(1,1)\over U(1)}\times {SO(2,2)\over SO(2)\times SO(2)}\times {SO(2,k_+)\over SO(2)\times SO(k_+)}\times {SO(2,k_-)\over SO(2)\times SO(k_-)}\, ,
\end{equation}
whose complex dimension is $1+2+k_++k_-$. In these cosets, $k_+=+\infty$ is the number of real scalars $m_4=-n_4=-Q=+1$ with $m_5$ or $n_5$ arbitrary. Similarly, $k_-=+\infty$ is the number of ``anti-tachyons''  $m_4=-n_4=-Q=-1$ with $-m_5$ or $-n_5$. Due to the $\Z_2\times \Z_2$ orbifold action (see Eq.~\ref{orbifold}), we know that tachyons and anti-tachyons are identified. Among the coordinates $Z_a^S$, those which do not survive the truncation are  set to zero. For instance, the third coset is parameterized by $Z_a^S$, $a\in\{6,7\}$, where the superscript is restricted to $S\in\{12,\dots,11+k_+\}\equiv {\cal I}$,  and that satisfy $\sum_{S,T\in {\cal I}}\eta_{ST}Z_a^SZ_b^T=-\delta_{ab}$.

Once we know the supermultiplet content of the $\N=4$ supergravity, we need to  specify the gauging, i.e. the non-Abelian interactions between the gauge bosons belonging to the vector multiplets as well as the 6 graviphotons. This amounts to determining the structure constants $f_{RST}$, totally antisymmetric in their indices $R,S,T\in\{4,\dots,9+(2+k_++k_-)\}$. By supersymmetry, a potential is generated, which is~\cite{sugra2, sugra3,sugra4,sugra2bis,sugraN=4,GP2}
\begin{equation}
\label{poten}
V = {|\Phi|^2\over 4}\,  Z^{RU}Z^{SV}\Big( \eta^{TW} + \frac{2}{3}Z^{TW} \Big) f_{RST}f_{UVW}\, , 
\end{equation}
where $Z^{RU}=Z^R_aZ^U_a$. To understand how the structure constants can be determined, it is instructive to consider as an example the supersymmetric case ($e=0$), for which the left- and right-moving generalized momenta and squared mass for $m_4=-n_4=\epsilon$, $m_5=n_5=0$, $\vec Q^2=1$ take the following form:
\begin{equation}
p_{\stackrel{\scriptstyle4\rm L}{\;\;\rm R}}={\epsilon\over \sqrt{2}} \Big({1\over R_4}\mp R_4\Big)\, , \quad M^2= \Big({1\over R_4}- R_4\Big)^2.
\end{equation} 
When $R_4=1$, two vectors multiplets become massless and satisfy $p_{4\rm L}=0$, $p_{4\rm R}=\epsilon\sqrt{2}$. Recognizing $p_{4\rm R}$ to be the non-Cartan charges of $SU(2)$, one concludes that the massless vector multiplet enhance the $U(1)_{\rm L}\times U(1)_{\rm R}$ gauge symmetry generated by the dimensionally reduced metric and antisymmetric tensor, $(G+B)_{\mu 4}, (G-B)_{\mu 4}$, to $U(1)_{\rm L}\times SU(2)_{\rm R}$. As a result, in a supersymmetric string theory model at some given point in moduli space, the structure constants in a Weyl-Cartan basis are nothing but the generalized momenta evaluated in the associated background, $\langle p_{I\rm L}\rangle, \langle p_{I\rm R}\rangle$~\cite{GP2}. 

The generalization of this result when supersymmetry is spontaneously broken  ($e=1$) is not known. The main difficulty in that within a vector multiplet, the values of the generalized momenta depend on $\vec Q$.  However, because in our case of interest all scalar superpartners of the possible tachyons have masses of order $\Ms$, they can be safely set to zero and the potential $V$ can be expressed only in terms of the structure constants associated with the generalized momenta of the tachyonic modes. Labelling the latter by an index $A$ or $\bar A$, 
\begin{eqnarray}
&&A \equiv (m_4=-n_4=-Q=+1,\phantom{-}m_5,0) \mbox{ or } (m_4=-n_4=-Q=+1,0, \phantom{-}n_5)\, , \nonumber \\
&&\bar A \equiv (m_4=-n_4=-Q=-1,- m_5,0) \mbox{ or } (m_4=-n_4=-Q=-1,0, - n_5)\, ,  \nonumber
\end{eqnarray}
the non-trivial structure constants involving vector multiplets are, up to antisymmetry, 
\begin{eqnarray}
&&f_{4A\bar A} = \langle p_{4L}\rangle={1\over\sqrt{2}}\Big({1\over 2 \langle R_4\rangle}-\langle R_4\rangle\Big)  , \;  f_{10A\bar A} = \langle p_{4R}\rangle={1\over\sqrt{2}}\Big({1\over 2 \langle R_4\rangle}+\langle R_4\rangle\Big)\nonumber \\
&&f_{5A\bar A}= \langle p_{5L}\rangle={m_5\over \sqrt{2}\, \langle R_5\rangle} \;\;\mbox{ or } \;\;\;\;\;\;{n_5\over \sqrt{2}}\, \langle R_5\rangle\, ,\label{st1} \\
&& f_{11A\bar A}\! = \!\langle p_{5R}\rangle={m_5\over \sqrt{2}\, \langle R_5\rangle} \;\;\mbox{ or }\;\; -{n_5\over \sqrt{2}}\, \langle R_5\rangle\, .\nonumber
\end{eqnarray}
Moreover, the non-Abelian structure of the 6 graviphotons of $\N=4$ supergravity must be specified. For this purpose, we consider an ansatz consistent with the $\Z_2\times \Z_2$ orbifold action,
\begin{equation}
\label{st2}
f_{468}=e_{\rm L},\quad \;\;f_{10,68}=e_{\rm R},\quad \;\;f_{479}=\tilde e_{\rm L},\;\;\quad f_{10,79}=\tilde e_{\rm R}\, , 
\end{equation}
where the right hand sides will be determined by imposing the no-scale supergravity phase to reproduce data of the heterotic model. 


\section{Tree level potential}
\label{IV}

We are ready to derive the potential of the $\Z_2\times \Z_2$ truncated $\N=4$ supergravity, by using all ingredients introduced in the previous sections. In Eq.~(\ref{cosets}), the last three cosets can be reparameterized in terms of ``constrained'' variables $\phi^S$ satisfying $Z^{ST}=4\big(\phi^S\bar \phi^T+\bar \phi^S\phi^T\big)$. In particular, for the second manifold, we define
\begin{eqnarray}
&&\phi^4={1-TU\over \sqrt{y}}, \quad \phi^5={T+U\over \sqrt{y}},\quad \phi^{10}={1+TU\over \sqrt{y}}, \quad \phi^{11}={T-U\over \sqrt{y}},\nonumber \\
&&y=-(T-\bar T)(U-\bar U)>0\, ,
\end{eqnarray}
where $T,U$ are ``unconstrained'' complex coordinates. Similarly, for the third coset, we take 
\begin{eqnarray}
&&\phi^{6}={1\over 2\sqrt{Y}}\Big(1+\sum_{A}\omega_A\Big)\; , \quad \phi^{7}={i\over 2\sqrt{Y}}\Big(1-\sum_A\omega_A\Big)\; , \quad \phi^{A}={\omega_A\over \sqrt{Y}}\, ,\nonumber \\
&&Y\equiv1-2\sum_{A}|\omega_A|^2+\Big|\sum_{A}\omega^2_A\Big|^2>0\, ,
\end{eqnarray}
in terms of unconstrained  Calabi-Vesentini complex coordinates $\omega_A$. Finally, $\phi^8,\phi^9,\phi^{\bar A}$ can be expressed in terms of unconstrained coordinates $\omega_{\bar A}$ of the fourth manifold.

In order to identify tachyons and anti-tachyons, and to set to zero their massive superpartners (the tachyons belong to chiral multiplets), we impose $\omega\equiv \omega_{\bar A}\in \mathbb{R}$. Moreover, because we restrict our analysis to the case where the compact directions $X^4,X^5$ are factorized circles $S^1(R_4)\times S^1(R_5)$, we take the supergravity variables $T,U$ to be of the form $T=i{\cal R}_4{\cal R}_5$, $U=i{\cal R}_4/{\cal R}_5$. In these conditions, the truncated gauged $\N=4$ supergravity potential takes the following form~\cite{PV}
\begin{equation}
V={|\Phi|^2\over 2}\Big(C^{(0)}+C^{(2)}_A\Omega_A^2+C^{(4)}_{AB}\Omega_A^2\Omega_B^2\Big),\quad \;\; \Omega_A\equiv {\omega_A\over \sqrt{Y}}\, ,
\end{equation}
where $C^{(0)}, C_A^{(2)}, C_{AB}^{(4)}$ are explicitly given in terms of the moduli ${\cal R}_4, {\cal R}_5$ and the structure constants of Eqs~(\ref{st1}),~(\ref{st2}). The dictionary between the supergravity variables and the string theory moduli may not be trivial. Therefore, we introduce real coefficients $\gamma_{\rm dil}, \gamma_4,\gamma_5$ such that 
\begin{equation}
|\Phi|^2=\gamma_{\rm dil} \, e^{2\phi_{\rm dil}},\quad \;\;{\cal R}_4=\gamma_4 R_4,\quad\;\; {\cal R}_5=\gamma_5 R_5\, , 
\end{equation}
where $\phi_{\rm dil}$ is the string theory dilaton field. 
Imposing that in the no-scale supergravity phase, where all $\Omega_A$'s vanish,  the cosmological constant is zero, and the mass spectrum matches  Eq.~(\ref{mass}), we find two solutions ($\sigma=\pm 1$)
\begin{eqnarray}
&&e_{\rm L}=\langle p_{4\rm L}+\sigma \sqrt{3}p_{4\rm R}\rangle\, ,\;\;  \quad e_{\rm R}=\langle p_{4\rm R}+\sigma \sqrt{3}p_{4\rm L}\rangle\, ,\quad\;\; -\tilde e_{\rm L}^2+\tilde e_{\rm R}^2=2\,,\nonumber \\
&&\gamma_{\rm dil}={1\over 2}\, , \quad \;\;\gamma_4={2+\sigma \sqrt{3}\over \langle R_4\rangle}\, ,\quad\;\; \gamma_5={1\over \langle R_5\rangle}\, .
\end{eqnarray}
In the end, written  in terms of the heterotic string theory moduli fields, the potential takes the final form,
\begin{eqnarray}
V=e^{2\phi_{\rm dil}}\, 4\,\bigg\{\!\!\!\!&&\Big({1\over 4R_4^2}+R_4^2-3\Big)\sum_A\Omega_A^2+ {1\over R_5^2}\sum_{m_5}m_5^2\, \Omega_{A}^2+R_5^2\sum_{n_5}n_5^2\, \Omega_{A}^2\nonumber \\
&&+\Big({1\over R_4^2}+4R_4^2\Big)\Big(\sum_A\Omega_A^2\Big)^2\label{Vfin}\\
&& + {4\over R_5^2}\Big(\sum_{m_5}m_5\, \Omega_{A}^2\Big)^2+4R_5^2\Big(\sum_{n_5}n_5\, \Omega_{A}^2\Big)^2\bigg\}.\nonumber 
\end{eqnarray}

Some remarks are in order. First, we note that the duality transformations $R_4\to 1/(2R_4)$ and $R_4\to 1/R_5$, which are satisfied by the 1-loop heterotic string partition function, remain valid off-shell, at least at the low energy level, since they are symmetries of $V$ (as well as of the full effective action). Therefore, for the definition of the supersymmetry breaking scale to be valid for arbitrary $R_4$, we take
\begin{eqnarray}
\Msusy\equiv {1\over \sqrt{2}\, e^{|\ln (\sqrt{2}R_4)| }}\, .
\end{eqnarray}
Second, when the background value $\langle R_4\rangle$ sits outside the range given in Eq.~(\ref{range}), because all mass terms in the first line of Eq.~(\ref{Vfin}) are positive, it is clear that the no-scale phase of the heterotic model is recovered, with its degenerate vacua and flat directions:
\begin{equation}
\langle V\rangle=0\, , \quad \langle \Omega_A\rangle=0\, , \, \forall A\, , \quad\langle \Msusy\rangle  < M_{\rm c}\, , \quad\langle R_5\rangle,\, \phi_{\rm dil}\mbox{ arbitrary}\, .
\end{equation}
Third, when $\langle R_4\rangle$ sits in the range of Eq.~(\ref{range}), one finds two degenerate branches of extrema with respect to the $\Omega_A$'s and the radii:
 \begin{equation}
\langle\Omega_A\rangle=\pm {1\over 2}\, \delta_{m_5,0}\, \delta_{n_5,0}\, ,\quad \langle R_4\rangle ={1\over \sqrt{2}}\, ,  \quad \langle R_5\rangle \mbox{ arbitrary}\, .
\end{equation}
Only one scalar condenses, which is the tachyon with trivial momentum and winding numbers in all directions other than the Scherk-Schwarz circle $S^1(R_4)$. Expanding the condensing mode as $\pm 1/2+\delta \Omega_0$, and the radius as $R_4=1/\sqrt{2}+\delta R_4$, the potential becomes for small fluctuations of the fields
\begin{eqnarray}
V=e^{2\phi_{\rm dil}}\Big( \!-1+8\delta R_4^2+16\delta\Omega_{0}^2&&+ \frac{4}{R_5^2}\sum_{m_5}m_5^2\, \Omega_{A}^2\nonumber 
\\&&+4R_5^2\sum_{n_5}n_5^2\, \Omega_{A}^2 + \cdots\Big)\, .
\end{eqnarray} 
Therefore, $\delta R_4$, $\delta \Omega_0$ and all non-condensing $\Omega_A$'s are massive, while $R_5$ is massless. However, the dilaton field has a tadpole and cannot be stabilized. Actually, writing the effective action in string frame, $\hat g_{\mu\nu}=e^{2\phi_{\rm dil}}g_{\mu\nu}$, where $g_{\mu\nu}$ is the Einstein frame metric, one obtains
\begin{equation}
S_{\rm tree}=\int d^4x \sqrt{-\hat g}\,  e^{-2\phi_{\rm dil}}\Big({\hat {\cal R}\over 2}+2(\partial \phi_{\rm dil})^2+1+\O(\delta)+\mbox{other fields}\Big),
\end{equation} 
where $\hat {\cal R}$ is the Ricci curvature. 
Notice that this expression matches the action of a non-critical string theory with linear dilaton background $\phi_{\rm dil}=\kappa_\mu X^\mu+\phi_0$, where $\kappa_\mu$ is a constant vector. As a result, it may be that the new phase arising from tachyon condensation, and which is characterized by a negative potential, is associated with a new fundamental heterotic string theory in non-critical dimension~\cite{PV,AK,ADK}. 


\section{Conclusion}
\label{V}

In this note, we have considered classical heterotic string backgrounds realizing the spontaneous breaking of $\N=1$ supersymmetry  in Minkowski spacetime, and we have shown that the scale $\Msusy$ cannot exceed some critical value $\Mc=\O(\Ms)$. We have restricted our analysis to the case where the condensing tachyon has vanishing momentum and winding numbers along the internal directions not involved in the Scherk-Schwarz breaking of supersymmetry. However, as can be seen from Eq.~\ref{orbifold}, another choice of discrete torsion in the model imply all potentially tachyonic states surviving the orbifold action to have non-trivial momentum or winding in these directions. It would be very interesting to apply our approach to this case, in order to find all different regions in moduli space corresponding to new string theory phases.

Another interesting generalization of our work would be to take into account all metric and antisymmetric tensor modui-dependence of the torus of coordinates $X^4,X^5$. In that case, the region in moduli space where the tachyon condensation takes place is  much more involved. 

As a conclusion, let us mention that because in the very early universe the supersymmetry breaking scale is naturally of the order of the string scale, the phenomenon described in the present work may yield an alternative paradigm  to inflation or bouncing cosmologies. 



\end{document}